\documentclass[manuscript]{aastex}

\usepackage{ulem}  

\newcommand{\ltsimeq}{\raisebox{-0.6ex}{$\,\stackrel
        {\raisebox{-.2ex}{$\textstyle <$}}{\sim}\,$}}

\newcommand{\gtsimeq}{\raisebox{-0.6ex}{$\,\stackrel
        {\raisebox{-.2ex}{$\textstyle >$}}{\sim}\,$}}

\slugcomment{AJ revised \today}

\received{}
\revised{17 May 2010}
\accepted{}

\shorttitle{Rectified Asteroid Albedos and Diameters}
\shortauthors{Ryan \& Woodward}

\begin{document}


\title{RECTIFIED ASTEROID ALBEDOS AND DIAMETERS FROM IRAS AND MSX PHOTOMETRY
CATALOGS}


\author{ERIN LEE RYAN, CHARLES E. WOODWARD}

\affil{Department of Astronomy, School of Physics and Astronomy, 116 
Church Street, S.~E., University of Minnesota, Minneapolis, MN 55455\\ 
\it{ryan@astro.umn.edu, chelsea@astro.umn.edu}} 

\begin{abstract}


Rectified diameters and albedo estimates of 1517 main belt asteroid 
selected from the IRAS and MSX asteroid photometry catalogues are derived 
from updated infrared thermal models, the Standard Thermal Model (STM) and 
the Near Earth Asteroid Thermal Model (NEATM), and Monte Carlo simulations, 
using new Minor Planet Center (MPC) compilations of absolute magnitudes 
(H-values) constrained by occultation and radar derived parameters. The 
NEATM approach produces a more robust estimate of albedos and diameters, 
yielding albedos of $p_{v}$(NEATM mean)$=0.081 \pm 0.064$. The asteroid 
beaming parameter ($\eta$) for the selected asteroids has a mean value of 
$1.07 \pm 0.27$, and the smooth distribution of $\eta$ suggests that this 
parameter is independent of asteroid properties such as composition.  
No trends in $\eta$ due to size-dependent rotation rates are evident. 
Comparison of derived $\eta$'s as a function of taxonomic type 
indicates the beaming parameter values for
S-type and C-type asteroids are identical within the standard deviation of 
the population of beaming parameters. 

\end{abstract}



\section{INTRODUCTION\label{intro}}

The study of planet building in protoplanetary disks is an emerging area of 
emphasis. The number of detected exoplanetary systems continues to increase 
\citep[e.g.,][]{bouchy09,otoole09} and the inventory of dust mineral 
composition, organic materials, water abundances, ices, and gas content in 
the planet forming regions is now routinely determined from remote sensing 
spectrophotometry obtained by the NASA \textit{Spitzer}  Space Telescope 
and other facilities. However, the process by which micron-sized dust 
grains, volatile ices, and gas coalesce, aggregate, and grow leading to 
large planet-sized bodies is not well understood. Within our solar system, 
the main belt and Near-Earth asteroids (NEAs), are relics from the epoch of 
planet-building \citep{bottke05}. These objects can be statistically 
surveyed to characterize the size-frequency distribution, related to the 
evolution of planetesimals the solar system, to determine the albedo 
distribution, a measure of compositional gradients extant at the epoch of 
planet building, and to constrain models of solar system evolution 
\citep[e.g.,][]{gomes05}. 

However, models of solar system formation are primarily constrained by 
albedo calculations from the Infrared Astronomical Satellite 
\citep[IRAS;\,][]{iras02} and the Mid-Course Satellite Experiment 
\citep[MSX;\,][]{msx02} surveys. In this paper we re-analyze the asteroid 
photometry from IRAS and MSX with the Standard Thermal Model and the more 
recent Near Earth Asteroid Thermal Model to derive revised albedo and 
diameter estimates for asteroids detected in these surveys. We compare the 
results from our new analysis to occultation and radar derived diameters in 
an effort to assess which thermal model approach is of higher fidelity and 
therefore more applicable in the analysis of large volume ecliptic asteroid 
surveys such as those conducted by the NASA Wide-Field Infrared Satellite 
Explorer mission \citep[WISE;\,][]{wise08}. 

\section{ARCHIVAL ANALYSIS\label{obs}}

\subsection{\textit{Asteroid Selection from Catalogs}}

The IRAS catalog was produced for asteroid sightings in IRAS survey 
data \citep{irascat}. The flux catalog consists of 12, 25, 60, and 
100~\micron{} photometric data of varying qualities for 9244 asteroid 
sightings. In this catalog, the data quality flags (DQF) range from 
values of 1 (low-quality) to 5 (high-quality), the latter DQF value 
denoting asteroids detected a minimum of twice in a given IRAS 
photometric bandpass. We selected all asteroids from the IRAS catalog 
with a DQF value of 3 or greater, and with reported fluxes in either three 
or four IRAS bandpass for analysis. These selection criteria produce a 
catalog of 5940 sightings of 1425 individual asteroids. 

The MSX sightings catalog \citep{msxcat} contains 325 individual 
sightings of asteroids. Data quality from MSX is not flagged; therefore, 
our MSX selection criterion accepts all asteroids for which fluxes in 
either three or four bands are reported for a given single sighting. 
This results in a dataset of 185 sightings of 92 individual asteroids. 
Table~\ref{table:csac_tab}, the criteria-selected asteroid catalog 
(CSAC), summarizes all asteroids extracted from the IRAS and MSX catalogs. 
The entire table is available in an one-line electronic compilation.

\subsection{\textit{Models}\label{sec:therm_mods}}

To determine the albedos and diameters of asteroids present in IRAS and 
MSX data, we have utilized both the Standard Thermal Model 
\citep[STM;][]{las89} and the Near Earth Asteroid Thermal Model 
\citep[NEATM;][]{har98}. We used these models to generate a best-fit 
spectral energy distribution (SED) from all flux measurements 
simultaneously obtained at a given epoch for each asteroid in our CSAC 
database. Our approach differs from the methodology employed by 
\citet{irascat,msxcat} where albedos and diameters were determined in 
each observation band and then averaged for the reporting of a ``bulk" 
albedo.  

The STM and the NEATM commonly used to obtain diameters and albedos of 
asteroids from mid-infrared (IR) photometry incorporate a variety of 
assumptions, the validity of which can be tested given extensive and 
reliable observational data. Briefly, we highlight the assumptions of each 
model with particular emphasis on how model assumptions influence the 
derivation of diameters and albedos. 

The STM \citep{las89} is the model used for the analysis of IRAS data by 
\citet{irascat}. This model assumes that: an asteroid is a perfect sphere; 
the asteroid temperature distribution is function of angular distance from 
the sub-solar point; the IR flux scales linearly in magnitudes with phase 
angle; and 
the thermal flux from an asteroid only arises from the illuminated 
``day-side'' of an asteroid. The STM also invokes use of a fixed parameter, 
$\eta$, referred to as the beaming parameter. The value of $\eta$ is chosen 
such that the diameters of asteroids Ceres and Pallas derived from IR 
observations of \citep{las89} are equivalent to those determined from 
stellar occultation experiments (i.e., direct measures). The $\eta$ 
parameterization accounts for surface roughness and thermal inertia, 
tacitly assuming that all asteroids have a similar surface characteristics, 
and presupposes that all compositions have the same thermal inertia. 

The thermal distribution assumed for the STM is defined as

\begin{equation}
T_{STM}(\Omega)= [ \frac{(1-A)S_\odot}{0.756 r_{h}^{2} \epsilon 
\sigma}]^{\frac{1}{4}} \times [cos (\Omega)]^{\frac{1}{4}} \label{eqn:stm_e}
\end{equation}

\noindent where the temperature, T is in Kelvin, A is the geometric Bond 
albedo, S$_{\odot}$ is the solar constant (W~m$^{-2}$), r$_{h}$ is the 
heliocentric distance (AU), $\epsilon$ is the emissivity of the object 
(assumed to be 0.9), $\sigma$ is the Stefan-Boltzmann constant, and 
$\Omega$ is the angular distance from the sub-solar point. 

Because the STM does not provide good fits to NEAs, \citet{har98} introduced 
the NEATM model. The NEATM assumes an asteroid is a perfect sphere, the 
temperature distribution on the asteroid is a function of both latitude 
and longitude, the asteroid has a spin axis perpendicular to the Sun-Earth 
plane, and only the ``day-side" of the asteroid contributes to the 
thermal flux. This model does not scale the thermal flux as a function 
of phase angle, rather it numerically calculates the actual thermal flux 
detected from the illuminated portion of a smooth sphere visible at a 
given phase angle. The NEATM also utilizes a beaming parameter ($\eta$), but 
rather than treating it as a fixed value the model defines $\eta$ as a 
measure of deviation from a smooth body with zero thermal inertia. As a 
free parameter, $\eta$ is best-fit simultaneously with the geometric 
albedo of a target. 

The object thermal distribution assumed for the NEATM is:

\begin{equation}
T_{NEATM}(\Omega)= [ \frac{(1-A)S_\odot}{\eta r_{h}^{2} \epsilon 
\sigma}]^{\frac{1}{4}}  (cos \phi)^{\frac{1}{4}} (cos \theta)^{\frac{1}{4}} 
\label{eqn:neatm_e}
\end{equation} 

\noindent where $\phi$ is the latitude, and $\theta$ is longitude. All 
other variables in this relation are defined as in Eqn.~\ref{eqn:stm_e}. 

Our implementation of the STM and the NEATM fits the observed photometric data 
using the downhill simplex method of chi-squared minimization 
\citep{nrc93}. The asteroid diameters are derived from the relation of 
\citet{fc92}, 

\begin{equation}
D(km)= \frac{1329}{\sqrt{p_{v}}}10^{\frac{\rm{-H}}{5}} \label{eqn:fc_dia}
\end{equation}

\noindent where $p_{v}$ is the geometric albedo and H is asteroid absolute 
magnitude for both models. Values for H are taken from the Minor Planet 
Center\footnote{www.cfa.harvard.edu/iau/mpc.html} (MPC). Many H-values have 
significantly changed and updates have been made to MPC compilations since 
the original dates of the IRAS and MSX publications (e.g., 
Table~\ref{table:csac_tab}). Those objects detected with signal-to-noise 
ratios $<10$ have their fluxes over predicted by the IRAS point source 
extraction routines \citep{iras02}; however, these flux overestimates are 
linear with respect to signal-to-noise ratio and can be corrected with a 
multiplicative factor. The point source extracted fluxes
(columns labeled ``flux''), the multiplicative
flux correction factor (columns labled ``flux Correction''), and 
overestimate corrected fluxes (columns labeled ``Corrected Flux'')
are reported in the CSAC. 

We also utilize the color corrections published for IRAS and MSX 
\citep{chas88, egan99}; these color corrections are calculated in 10 degree 
bins. We do not color correct the IRAS photometry based on the 
sub-solar temperature. Instead, we apply the color correction using the 
mean temperature of the illuminated face of the asteroid which allows the 
asteroid to have a non-zero thermal inertia (e.g., dependent on surface 
properties, rotational periods). The effective blackbody temperature of the 
asteroid SEDs utilizing the IRAS fluxes for a given sighting are cooler 
(i.e., the SEDs peak near 25~\micron) than that of the sub-solar temperature 
inferred for the effective heliocentric distance. Therefore, we calculate 
the mean temperature (from the temperature distribution defined in each 
model) as a proxy for the effective temperature of the object for the 
purposes of color correction. Use of the sub-solar temperature for the 
color correction is especially problematic for the NEATM fits to asteroid 
photometry, as the adoption of a sub-solar temperature for the color 
correction introduces a somewhat simplified first order assumption 
regarding the thermal inertia of the body (i.e., the body has zero thermal 
inertia such that it is in instantaneous equilibrium with the radiation 
field), and hence complicates interpretation of the derived values of 
$\eta$. Our approach is a departure from the \citet{irascat,msxcat} 
methodology results which utilize the sub-solar temperature for color 
corrections. 

Table~\ref{table:flux_color_fixed} reports example orbital parameters and 
fluxes utilized interatively in the thermal model fitting of 
IRAS data after: (1) flux overestimation corrections have been applied to
IRAS point source fluxes (Table~\ref{table:csac_tab}); and (2)~mean 
temperature analysis (an intermediate interative modelling step) 
color corrections are applied to the observed fluxes. Geometric 
albedoes and diameters derived from using the mean temperature 
color corrected fluxes in each model are also provided in this
table. Entries in Table~\ref{table:flux_color_fixed} should 
enable corroboration of our methodology.
The entire table is available in an one-line electronic compilation.

The $\chi^{2}$ minimization fitting of the asteroid photometric data from 
the STM and the NEATM do not yield formal errors for each fit parameter. To 
determine the uncertainty of the albedo, diameter, and where applicable 
$\eta$ derived from the models, a Monte Carlo code was utilized in 
conjunction with the observed photometry to create 100 synthetic asteroid 
sightings per sighting for each body in Table~\ref{table:csac_tab}. The 100 
synthetic sightings retained the orbital information such as heliocentric 
($r_{h}$) and geocentric distance ($\Delta$), but varied the photometric 
flux within the photometric uncertainties cited in each of the original 
source catalogs for a given asteroid. Results of our model fits with Monte 
Carlo simulation of the photometric uncertainty is presented in 
Table~\ref{table:mc_sum}. This table contains both the 
STM and the NEATM albedo 
and diameter determinations for each sighting. In the solutions catalog 
entries for which the STM solutions did not converge are marked with a 
value of -1000.0 in the albedo, albedo error, diameter, and diameter error 
columns. 

\section{RESULTS \& DISCUSSION\label{sec:disc}}

\begin{figure}
\epsscale{1.0}
\plotone{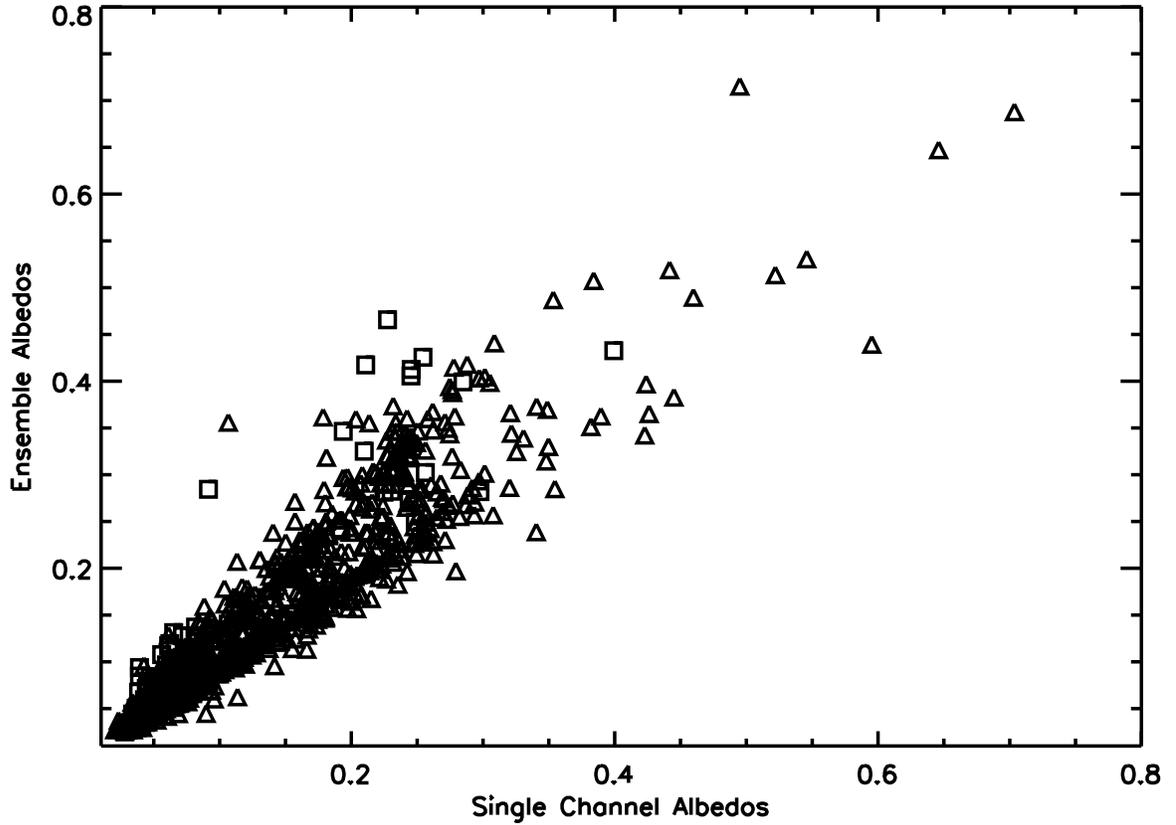}

\caption{Mean single channel albedos versus newly recalculated mean 
ensemble albedos for IRAS and MSX sources summarized in 
Table~\ref{table:csac_tab}. The \textit{open triangles} correspond to 
IRAS sources, while \textit{open squares} denote asteroids from the MSX 
catalogue. The correlation slope has a value of 1.10 for IRAS asteroids 
and 1.28 for MSX asteroids. 
\label{fig:ensem_v_single}}

\end{figure}

Mean albedo and diameter values for asteroids from MSX and IRAS are 
reported in our ancillary on-line catalog, whose format and example 
entries are given in Table~\ref{table:on_line_examp}. All subsequent
analysis, discussion, and conclusions in this manuscript are based 
on the data in Table~\ref{table:on_line_examp}. Comparison of our 
STM results with those in previously published studies 
\citep[e.g.,][]{irascat, msxcat}, indicate that the albedos obtained 
using an ensemble of four simultaneous measured photometric bands are 
systematically bluer than those determined from the mean of single 
channel (i.e., single photometric band) albedos.  The offset between the 
mean single channel albedos and simultaneous ensemble albedos is 12\% of 
as depicted in Fig.~\ref{fig:ensem_v_single}. 

\begin{figure}
\epsscale{1.0}
\plotone{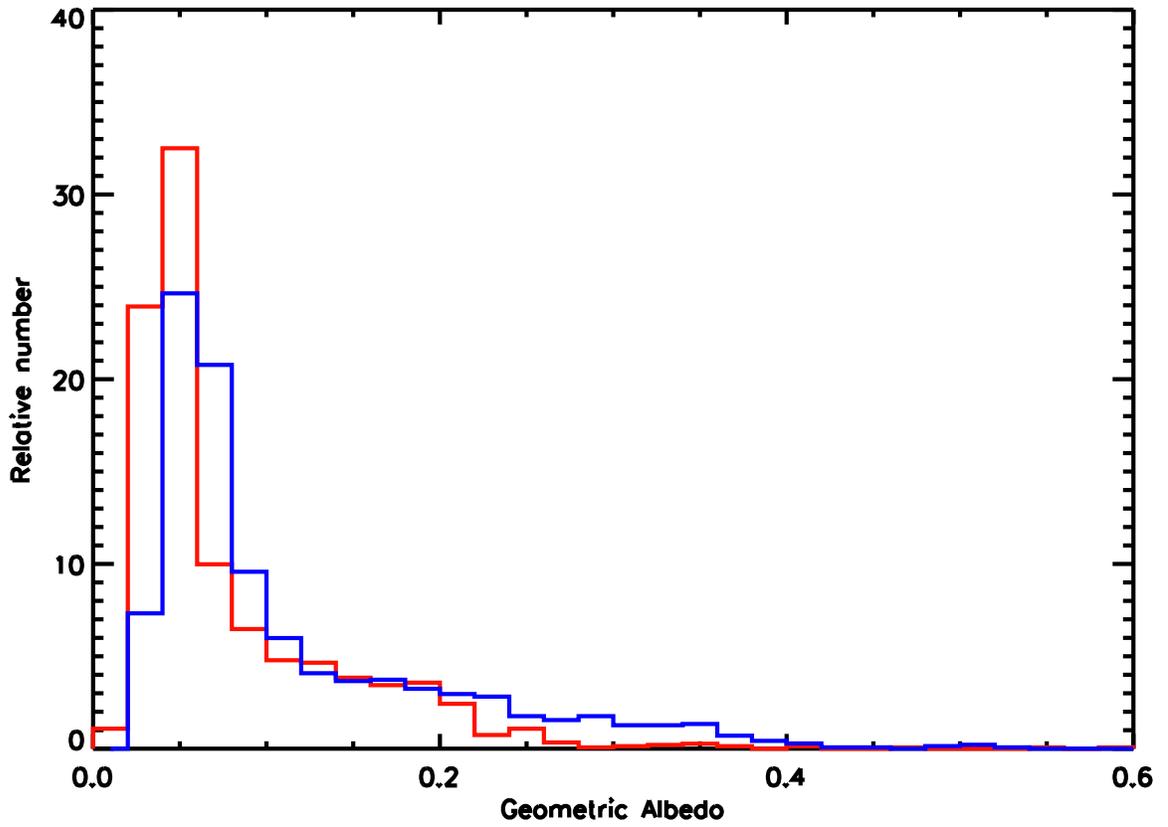}

\caption{Histogram distribution of the NEATM and STM infrared thermal 
model albedos derived for asteroids culled from the original IRAS 
asteroid photometry catalog. The red line denotes the NEATM results, the 
blue line denotes the STM albedos. 
\label{fig:hist_n_s}}

\end{figure}

The NEATM fits of asteroid albedo yield an albedo distribution for the main 
belt asteroids that are redder and narrower than the solutions from the STM. 
This effect is illustrated in Fig.~\ref{fig:hist_n_s}, a histogram 
distribution of the derived geometric albedos of each model. The mean 
albedo for asteroid NEATM fits is $p_{v}$(NEATM mean)$=0.081 \pm 0.064$ 
while the mean albedo for the STM solutions is $p_{v}$(STM mean)$=0.120 
\pm 0.099$. Canonically, the mean albedo adopted for main belt 
asteroids is 0.11 \citep{parker08,yoshida07,ivez01}. 

\begin{figure}
\epsscale{1.0}
\plotone{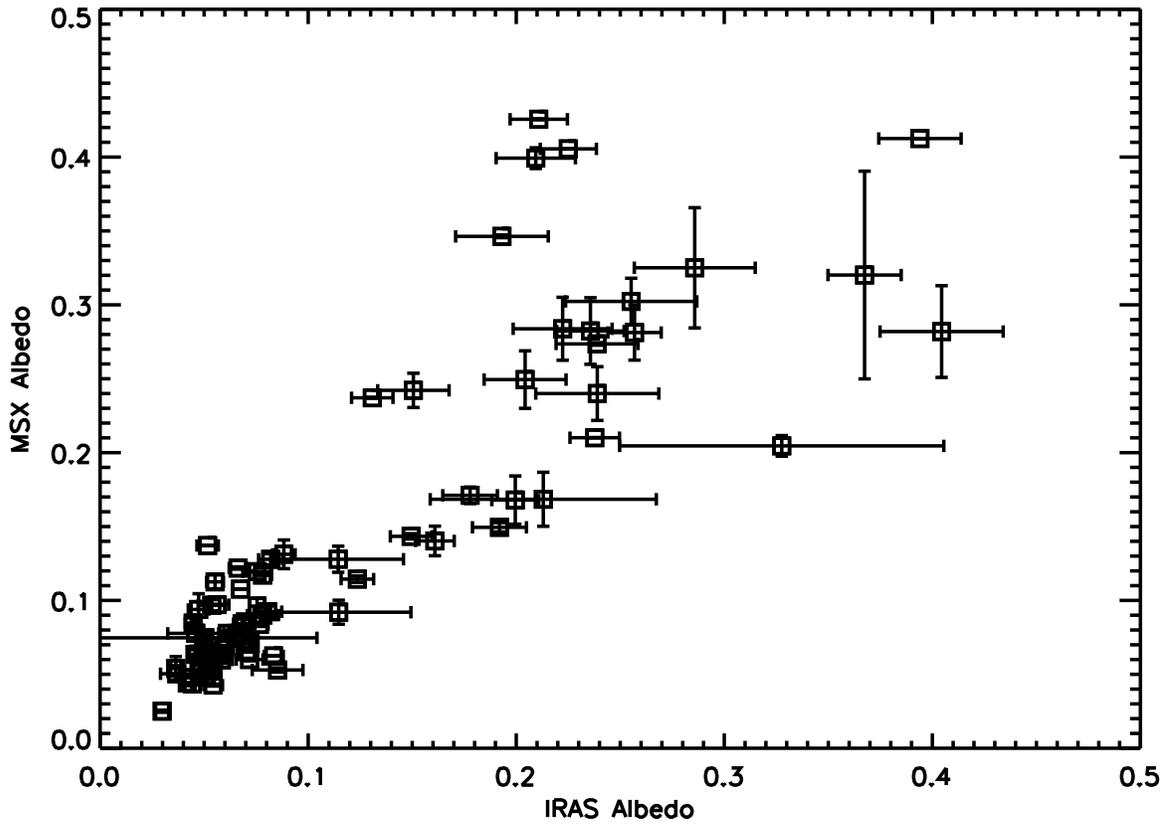}

\caption{Comparison of asteroid albedos derived from IRAS photometry 
compared to those derived from MSX photometry of the same objects. 
\label{fig:ivn_fig}} 

\end{figure}

Comparison of the albedos we have derived for individual objects cataloged 
in both the IRAS and MSX databases, to those estimated through use of 
either the IRAS or MSX observations alone differ with a dispersion of 
$\sim$ 4\%, Fig.~\ref{fig:ivn_fig}. Though many asteroids have multiple 
IRAS sightings, most of these sightings are separated by minutes. This 
observational cadence does not allow one to obtain infrared light curves 
for asteroids whose rotational periods have timescales of hours to days. The 
majority of asteroids sighted in the MSX catalog have photometry for only a 
single sighting, thus the photometry of these targets is not as robust as 
the observations from IRAS. Shape and inhomogeneous surface compositions as 
well as flux measurement errors could also account for the observed 
dispersion between the IRAS and MSX albedos. 

\subsection{\textit{Validation of Model Asteroid Diameters}\label{valid_mod}}

Independent best-fits SEDs generated using the STM and the NEATM to the 
identical asteroid photometric datasets can yield very disparate results 
for derived diameters and albedos. For example, the NEA 1627~Ivar was 
observed in the IR by \citet{delbo03}. The solutions for this target  using 
the STM and the NEATM yield geometric albedos ranging from 0.05 to 0.20, 
corresponding diameters between 9.12~km to 7.94~km respectively. The radar 
derived effective diameter \citep{ostro90} of Ivar is 8.5~km. 

The validity of thermal models fits to the IRAS photometric data of 
\citet{iras02} is difficult to assess as the model-estimated diameters of 
Ceres are not equivalent to the occultation-derived determinations. 
This discrepancy is problematic as the STM model was optimized with a 
beaming parameter $\eta$ set to 0.756 for the large asteroids Ceres and 
Pallas which may have thermal characteristics that are distinct from the 
general asteroid population \citep{las89}. The derived diameters for the 
largest asteroids observed by IRAS apparently are systematically low 
due to the use of ``band-to-band'' albedo corrections \citep{iras02}. 
The band-to-band correction is a multiplicative factor of 1.12 
which is applied to 25 and 60~\micron{} derived albedos such that their 
values come into agreement with those albedos derived from 
12~\micron{} photometry. 

Derivation of the 
band-to-band correction is not clearly explained in either \citet{iras02}, 
\citet{tedesco92}, or \citet{tedesco94}, hence examination of the 
underlying assumptions (e.g., wavelength dependent $\eta$) and their 
overall validity is difficult. However, to test the general applicability 
of the \cite{iras02} invoked band-to-band corrections, we ran the STM on 
the single channel photometry at 12, 25 and 60~\micron{} and determined the 
mean difference between occultation and radar derived diameters (many of 
which were not earlier extant) to those derived from IRAS data. In this 
analysis we utilized occultation and radar derived diameters of 80 
asteroids and find the mean difference between the occultation-derived 
diameters and the 12~\micron{} only diameters was $\simeq 17$\%, and 
$\simeq 25$\% and $\simeq 5$\% for the 25 and 60~\micron{} only diameters, 
respectively. This analysis suggests that use of a single channel albedo 
correction has highest fidelity when applied to IRAS 60~\micron{} 
photometry. 

\begin{figure}
\epsscale{1.0}
\plotone{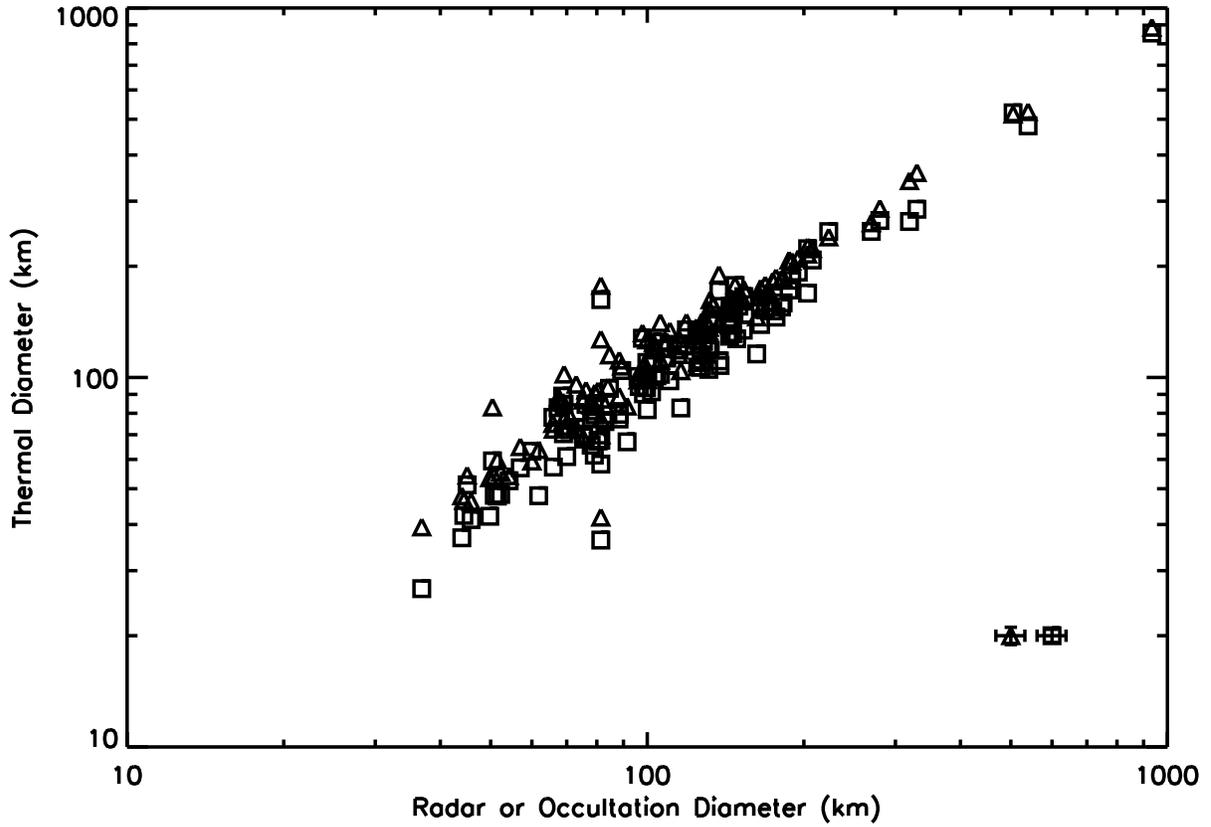}

\caption{Asteroid diameters estimated from infrared thermal models versus 
diameters derived from either radar or stellar occultation experiments. 
\textit{Open squares} denote diameters from STM calculations, while the 
\textit{open triangles} denote diameters computed from NEATM models. The 
insert two symbols in the bottom right hand corner of the panel are 
indicative of the average diameter uncertainties for the thermal model 
diameters. The mean error for radar determined diameters is of order 6.4\%. 
\label{fig:mods_v_ro_fig}} 

\end{figure}

\begin{figure}
\epsscale{1.0}
\plotone{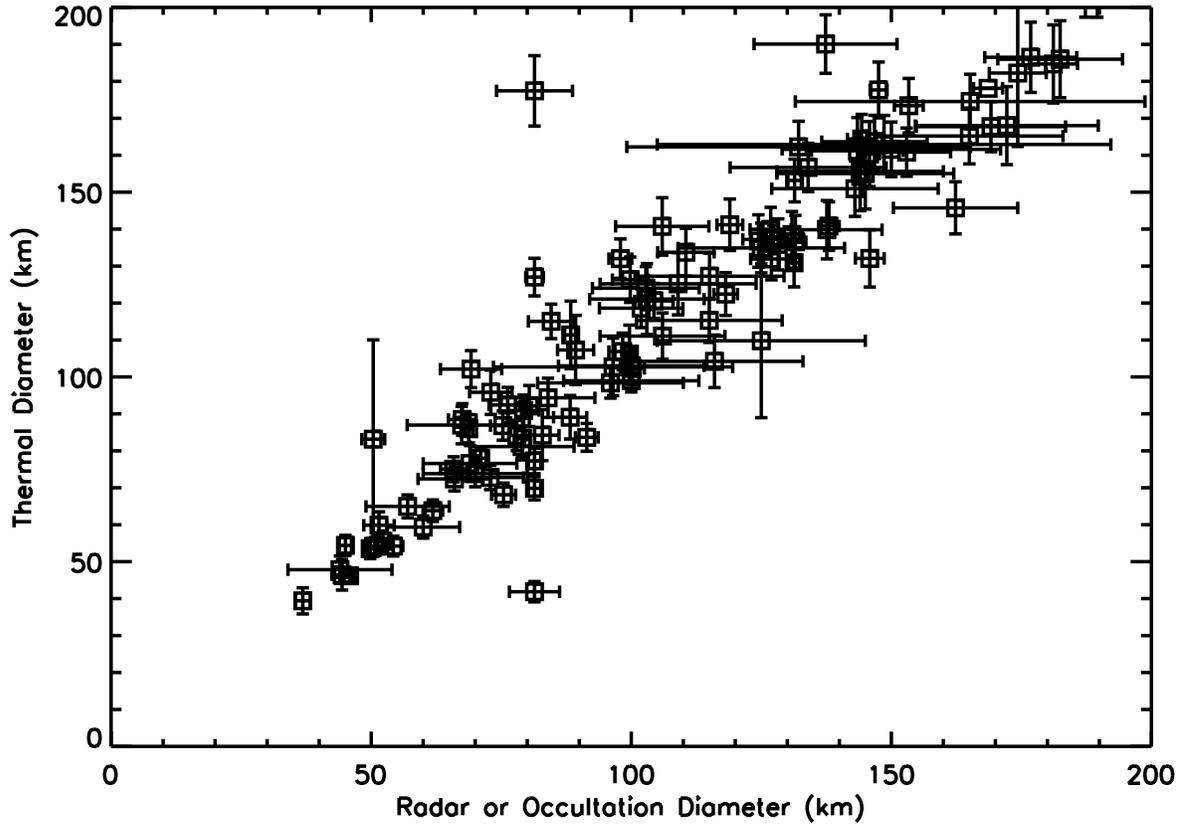}

\caption{Correlation of NEATM derived asteroid diameters versus radar or occultation 
diameters.
\label{fig:neatm_radar_fig}}

\end{figure}

To determine which thermal model (either the NEATM or the 
STM) approach generates 
diameters that are commensurate with asteroid diameters determined by 
independent and/or direct measurement techniques, we have compared object 
diameters in Table~\ref{table:on_line_examp} to those diameters established 
by radar or occultation observations. Cross-correlation of catalog entries 
in the IRAS, radar, and occultation databases yields 118 asteroids for this 
inter-comparison, Table~\ref{table:adtro}. Generally, the STM 
underestimates asteroid diameters by $\sim 10$\% when compared to radar and 
occultation derived estimates, while the NEATM underestimates diameters by 
$\sim 4$\% as illustrated in Fig.~\ref{fig:mods_v_ro_fig}. In absolute 
terms, application of the NEATM approach yields refined diameters 
commensurate with radar and occultation measurements, 
Fig.~\ref{fig:neatm_radar_fig}. However, there are moderate uncertainties 
($\simeq 10$\%) in the formal error of the derived diameters quoted for 
many asteroids observed with either radar or occultation technique. Thus, 
application of either the STM or the NEATM approach to model IR asteroid 
photometry is reasonable with appropriate assumptions and caveats. 

\subsection{\textit{The beaming parameter, $\eta$}\label{eta_disc}}

\begin{figure}
\epsscale{1.0}
\plotone{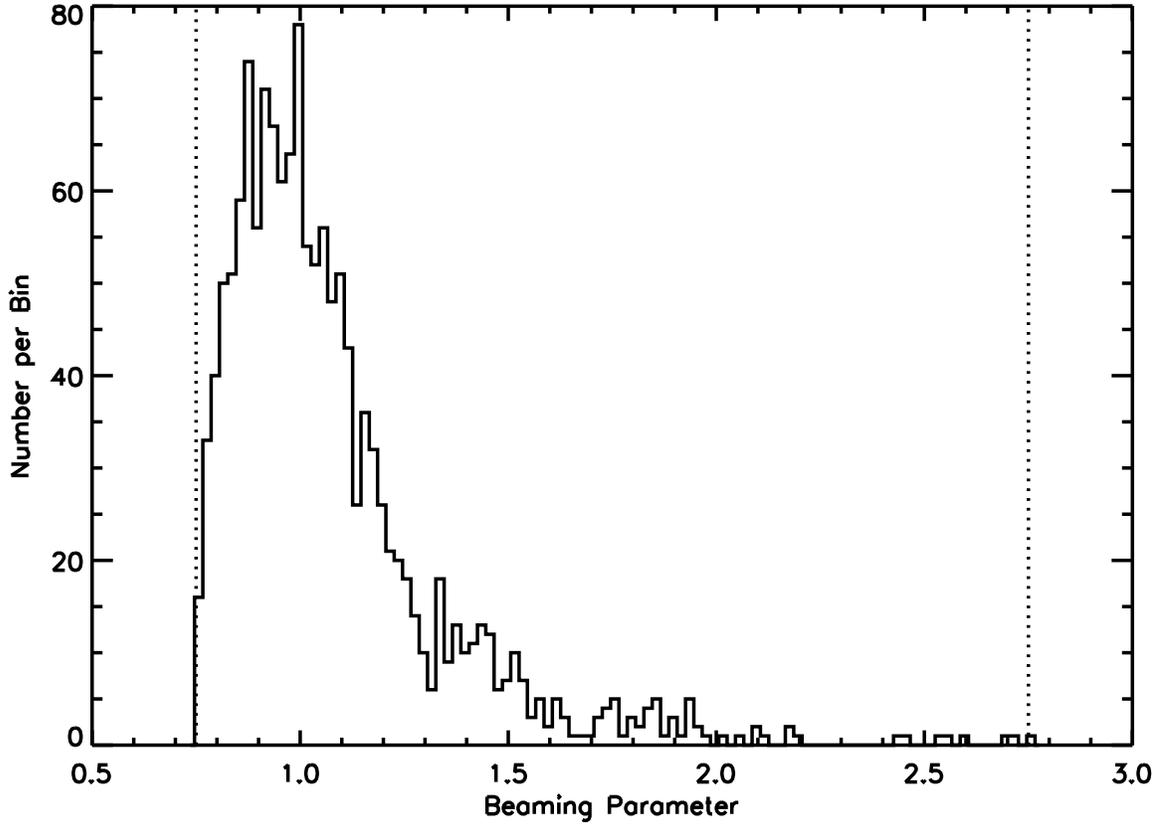}

\caption{Histogram distribution of the derived beaming parameter, $\eta$, 
of main-belt asteroids from NEATM best-fit models to the spectral energy 
distribution assembled from IRAS and MSX photometry. The vertical dashed
lines denoted limiting values set in the analysis code to force convergence.
\label{fig:eta_fig}} 

\end{figure}

The beaming parameter, $\eta$, within the main belt asteroids follows an 
approximate Gaussian distribution for the 1481 bodies in our sample 
population, Fig.~\ref{fig:eta_fig}. The cut-off in the histogram 
distribution at values of $\eta = 0.75$ and $\eta = 2.75$ are artifacts 
due to truncation criteria set within our code. The mean $\eta$ for 
asteroids is $1.07 \pm 0.27$ (exclusive of those within the histogram 
bins at 0.75 or 2.75). The smooth distribution of $\eta$ suggests that 
the beaming parameter is independent of asteroid properties such as 
composition. Comparison of the derived asteroid beaming parameters as a 
function of \cite{demeo09} taxonomic type indicates that $\eta$ for 
S-type and C-type asteroids are identical within the standard deviation 
of the population beaming parameters. 

\begin{figure}
\epsscale{1.0}
\plotone{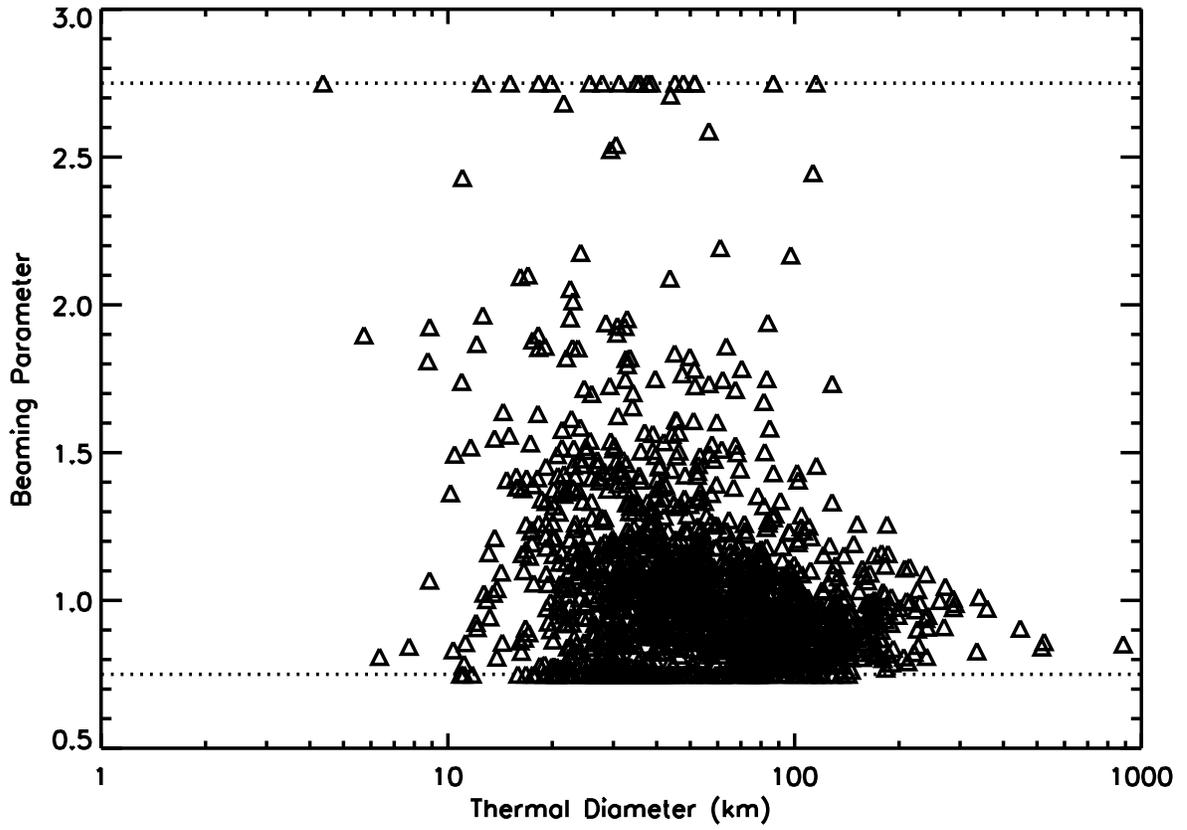}

\caption{Derived beaming parameter, $\eta$, as a function asteroid 
diameter. The horizontal dotted lines are limits imposed in the modeling 
code to force convergence. 
\label{fig:eta_diam_fig}} 

\end{figure}

\citet{delbtan09} assert, based on recent analysis of IRAS asteroid 
photometry, that the thermal inertia behavior of asteroids is a 
power-law function of diameter, with a break in the relationship 
occurring at size of $\sim 100$~km. \citet{delbtan09} speculate that this 
change in functional behavior may be due to differing regolith properties 
of small versus large asteroids. If regolith properties are size-dependent, 
trends in model-derived $\eta$ values should be apparent on our analysis. 
Figure~\ref{fig:eta_diam_fig} presents the beaming parameter fits as a 
function of asteroid diameter derived from the NEATM modeling. For 
diameters $\gtsimeq$ 100~km, $\eta$ has a fairly narrow range of values, 
$1.5 \ltsimeq \eta \ltsimeq 0.8$. Asteroids with smaller diameters have a 
much greater range of $\eta$, and only asteroids with diameters less than 
$\sim 100$~km are values of $\eta$ significantly greater than 1.6 
frequently apparent. The latter trend is consistent with the thermal 
inertia behavior suggested by \citet{delbtan09}.  

\begin{figure}
\epsscale{1.0}
\plotone{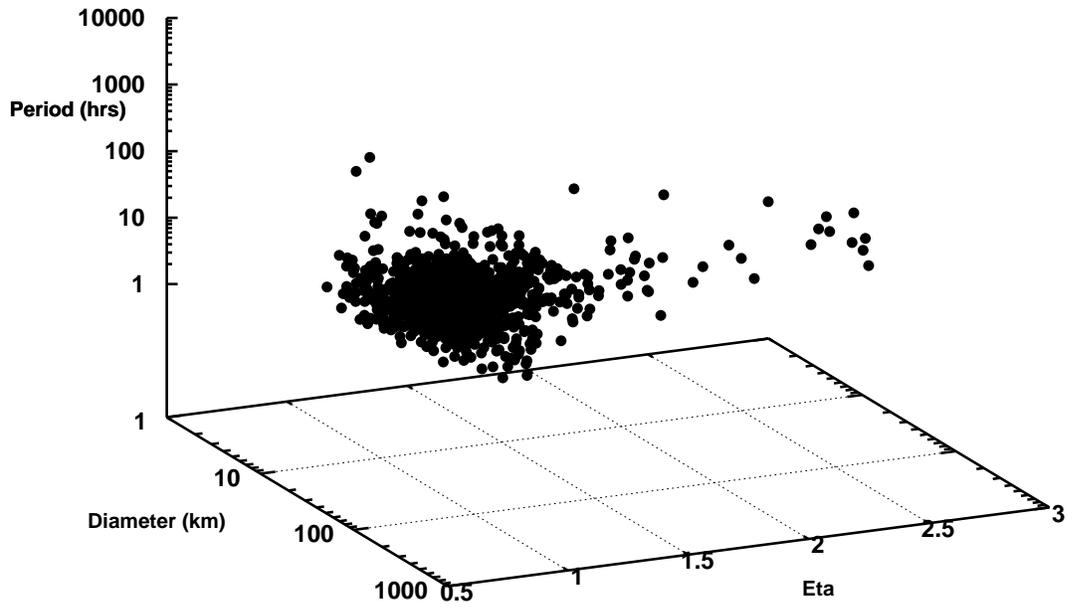}

\caption{The rotational period, diameter, $\eta$ phase-space for 
all asteroids within the study sample (Table~\ref{table:adtro})
with known rotational rates as published in the catalog of \citet{harris08}. 
\label{fig:diam_eta_period} } 

\end{figure}

Whether or not the behavior of $\eta$ as a function of asteroid 
diameter is solely dependent on size and therefore unique requires that 
affects of size-dependent rotation rates are minimal. Faster rotational 
periods also can give rise to increased $\eta$ depending on the pole 
orientation. Figure~\ref{fig:diam_eta_period} presents the period, diameter, 
$\eta$ surface for all asteroids in our survey which have tabulated 
rotational periods in the catalog of \citet{harris08}. There are no 
systematic trends or correlations between $\eta$ values and rotation 
periods for a given diameter. Asteroids with high $\eta$ values ($> 2.0$) 
are not required to be monoliths rather that rubble piles. 

The diameter discrepancies between the STM and the 
NEATM confirm that use of a fixed beaming
parameter $\eta = 0.756$ in the STM is not valid, especially for 
those asteroids with diameters $<$ 100 km.
The diameters obtained by \citet{delbtan09} are commensurate with those we 
have derived using the NEATM (Table~\ref{table:sn_v_delbo}). Hence the 
NEATM appears to be the preferable 
model to invoke to derive diameters from photometry. 

\subsection{\textit{Sparse-Sampling Effects}\label{sparse_disc}}

Our thermal model analysis of asteroid SEDs derived from simultaneous 
multi-band photometry enables quantitative assessment of the relative 
quality of albedo and diameter determined for asteroids that have only 
single photometric band (i.e., single-channel) measurements. Comparison of 
albedos derived solely from 12~\micron{} photometry to those resulting from 
model fits to SEDs differ by $\simeq 6$\%, while the variance between 
25~\micron{} only albedo fits and SEDs estimates are $\ltsimeq 45$\%. This 
discrepancy between albedo solutions may account for the distinct 
differences in our derived diameters and albedos 
(Table~\ref{table:on_line_examp}) compared to those previously published in 
the literature \citep{irascat, msxcat} even though a band-to-band
correction of the order 12\% was applied to the 25 and 60~\micron{} derived 
albedos from \citet{iras02}. However, the 12~\micron{} only albedo fits 
are in better consonance with those derived from the ensemble SED 
model-fits because the 12~\micron{} bandpass from IRAS is narrower than the 
25~\micron{} bandpass. Although color corrections exist for all IRAS 
wavebands, constraining albedo by use of the 25~\micron{} photometry is 
difficult as the bandpass is $\simeq$ 15~\micron{} wide, and many albedo 
solutions can fit this single photometric point. 

Asteroid albedos derived from MSX photometry show no clear divergence 
between the albedos determined from 4-channel simultaneous photometry (SED 
modeling) and albedos obtained using the 12 and 14~\micron{} channels 
individually. However, use of either the shortest or longest wavelength 
photometry does produce discrepant albedo estimates: the 21~\micron{} 
photometry  yields a variance of $\approx$~10\%, while the 8~\micron{} 
produces an incongruity of $\approx$~54\% compared to the SED derived 
values. The large differences in the derived albedos using the 8~\micron{} 
photometry is most likely due to solution ambiguities arising from the 
steepness of the greybody curve on the Wien side of the Planck flux 
distribution and the 20\% photometric uncertainty of the reported 
8~\micron{} flux densities. 

\subsection{\textit{Spectroscopic Type and Albedo}\label{sta_disc}}

Future all sky surveys, such as currently being conducted by the WISE 
mission or those planned by Pan-STARRS \citep{veres09} or the Large 
Synoptic Survey Telescope \citep{jones09}, require a means to 
relate asteroid colors to albedos in order to assess the NEA size, 
population, and earth impact threat. In the optical, the precise 
correlation of albedo and (B-V) color is uncertain. \citet{veeted92} find two 
loci in albedo vs. (B-V) color, one clustering near (albedo, B-V) = 
($\approx 0.05, \approx 0.7$), the other near ($\approx 0.22, \approx 
0.85$) which we can reproduce from asteroids albedos derived from our NEATM 
models and their corresponding (B-V) colors derived from catalog entries 
curated  at the Planetary Data System\footnote{http://pds.nasa.gov/} 
node \citep{lag95}. To break this degeneracy many of the 
newer asteroid surveys are expanding into the 
near-IR \citep[e.g.,][]{trilling09, ryan09}. The Bus-DeMeo 
taxonomy \citep{demeo09} is the first system which extends asteroid 
taxonomy to the IR (out to 2~\micron). 

\begin{figure}
\epsscale{1.0}
\plotone{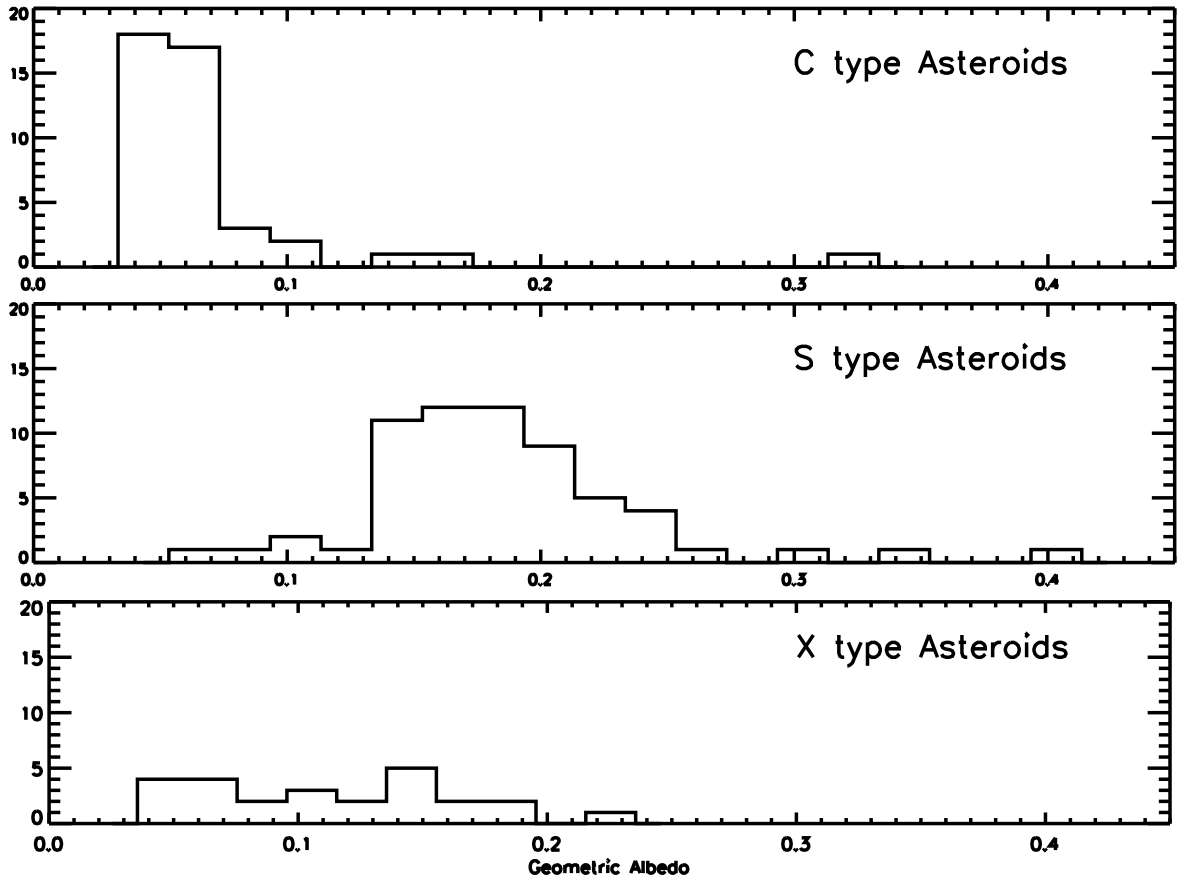}

\caption{Histograms of asteroid NEATM albedos by taxonomic type, as 
defined by \citet{demeo09}, for C-, S- and X-type classified 
asteroids in the study sample.
\label{fig:taxonomy_albedos} }

\end{figure}

Inspection of the NEATM albedos for the Bus-DeMeo taxonomic types 
(Table~\ref{table:demotax}) indicates that the
S- and C-type complexes have relatively narrow ranges of albedo. The NEATM 
albedo histogram distributions of asteroids with known Bus-DeMeo S-, C- and 
X-type classifications is presented in Fig.~\ref{fig:taxonomy_albedos}. 
The NEATM albedo values for these classes are lower in value than 
the STM derived albedos. The NEATM albedo range best-matches the 
spectroscopically inferred 
compositions of S- and C-class asteroids as suggested by laboratory 
albedo measurements of ordinary chondrite and carbonaceous chondrite 
materials \citep{piironen98}. Possibly, observations of asteroids in 
photometric bands akin to the Sloan Digital Sky Survey 
\citep[SDSS;][]{parker08} filter set, when extended into the near-IR, could  
provide a basic taxonomic classification of a newly discovered object 
reliable enough to obtain an asteroid diameter within $\sim 10$\%.  
However, the veracity of this assertion cannot be assessed as asteroids 
selected for our thermal analysis (Table~\ref{table:csac_tab}) are 
sufficiently large enough in size that they would have saturated the SDSS 
images and are thus not present in their moving target catalog 
\citep{sdss02}. 
       
\section{CONCLUSIONS\label{concl}}


We have derived new asteroid diameters and albedo estimates utilizing 1517 
objects selected from the IRAS and MSX asteroid catalogues using updated 
infrared thermal models, the STM and the NEATM. Spectral 
energy distributions using 
multi-band simultaneous infrared photometry and new values for asteroid 
absolute magnitudes (H-values) compiled in the MPC were fit and rectified 
using available occultation and radar derived diameters as constraints. Our 
model analysis suggests that the NEATM produces a more robust estimate of 
albedos and diameters. With the NEATM approach we find that the mean 
asteroid albedo is $p_{v}$(NEATM mean)$=0.081 \pm 0.064$, suggesting that 
the canonical albedo adopted for main belt asteroids,
0.11,\citep{parker08,yoshida07,ivez01} may be an overestimate. The mean 
beaming parameter, $\eta$, of asteroids in our select compilation is $1.07 
\pm 0.27$. The smooth distribution of $\eta$ suggests that the beaming 
parameter is independent of asteroid properties such as composition
and no trends in $\eta$ due to size-dependent rotation rates are evident. 
Comparison of the derived asteroid beaming parameters as a function of 
\cite{demeo09} taxonomic type indicates that $\eta$ for S-type and C-type 
asteroids are identical within the standard deviation of the population 
beaming parameters.

\acknowledgements

E.L.R. and C.E.W acknowledge support for this work provided by the National 
Science Foundation through the grant AST-0706980 issued to the University 
of Minnesota. The authors also wish to thank the anonymous referee whose 
insight and probing queries greatly improved this manuscript. 

{\it Facilities:} \facility{IRAS}
                   \facility{MSX}

\clearpage


\clearpage

%


\begin{deluxetable}{lcccccccccccccccccccccccccr}
\rotate
\tabletypesize{\tiny}
\tablecaption{INPUT FLUXES FROM IRAS AND MSX\label{table:csac_tab}}
\setlength{\tabcolsep}{1pt}
\tablewidth{0pt}
\tablehead{
\colhead{Source} & \colhead{Number} & \colhead{$r_{h}$} & \colhead{$\Delta$} & \colhead{phase} & \colhead{H} & \colhead{G} & \colhead{wave1} & \colhead{flux1} & \colhead{flux1} & \colhead{flux1} &  \colhead{Corrected}&  \colhead{wave2} & \colhead{flux2} & \colhead{flux2} & \colhead{flux2} & \colhead{Corrected} & \colhead{wave3} & \colhead{flux3} & \colhead{flux3} & \colhead{flux3}  & \colhead{Corrected} & \colhead{wave4} & \colhead{flux4} & \colhead{flux4} & \colhead{flux4} & \colhead{Corrected}\\

\colhead{} & \colhead{} &\colhead{} & \colhead{} & \colhead{}& \colhead{}& \colhead{}& \colhead{}& \colhead{}&\colhead{Error} &\colhead{Correction} &\colhead{Flux 1} &\colhead{}& \colhead{}&\colhead{Error} &\colhead{Correction}& \colhead{Flux 2} & \colhead{}& \colhead{}&\colhead{Error} &\colhead{Correction}& \colhead{Flux 3} & \colhead{}& \colhead{}&\colhead{Error} &\colhead{Correction}  &\colhead{Flux 4}\\

\colhead{} & \colhead{} & \colhead{(AU)} & \colhead{(AU)} & \colhead{(deg)} & \colhead{(mag)} & \colhead{} & \colhead{($\mu$m)} & \colhead{(Jy)} &\colhead{(Jy)} &\colhead{} &\colhead{(Jy)}& \colhead{($\mu$m)} & \colhead{(Jy)} &\colhead{(Jy)} &\colhead{} &\colhead{(Jy)}& \colhead{($\mu$m)} & \colhead{(Jy)} &\colhead{(Jy)} &\colhead{} &\colhead{(Jy)}& \colhead{($\mu$m)} & \colhead{(Jy)} &\colhead{(Jy)} &\colhead{} &\colhead{(Jy)}//
}

\startdata

  IRAS     &  1  & 2.951 &  2.769 &   20.030 &    3.340 &    0.120 &   12.000 &  374.673 &   36.116 &    1.000&   374.673 &   25.000 &  642.159  &  70.440  &   1.000 &  642.159 &   60.000 &  333.461 &   59.195  &   1.000 &  333.461 &  100.000 &  119.630  &  20.849  &   1.000  & 119.630\\
  IRAS   &    1 &  2.951 &  2.768 &   20.030  &   3.340 &    0.120 &   12.000 &  312.981 &   30.129 &    1.000&   312.981 &   25.000  & 628.203 &   68.761 &    1.000 &  628.203  &  60.000 &  297.782 &   39.112  &   1.000  & 297.782 &  100.000 &  108.973 &   20.317 &    1.000 &  108.973\\
 \enddata
\end{deluxetable}
\clearpage

\begin{deluxetable}{lcccccccccccccccccc}
\rotate
\tablecaption{STM AND NEATM SOLUTIONS PER SIGHTING\tablenotemark{a}
\label{table:flux_color_fixed}}
\tabletypesize{\tiny}
\setlength{\tabcolsep}{1pt}
\tablewidth{0pt}

\tablehead{
\colhead{} & \colhead{} & \colhead{} &\colhead{} &\colhead{} & \colhead{} &\colhead{STM} & \colhead{color} & \colhead{ corrected} &\colhead{} & \colhead{STM} & \colhead{results} & \colhead{NEATM} & \colhead{Color} & \colhead{Corrected} & \colhead{} &\colhead{NEATM} & \colhead{results} & \colhead{}\\

\colhead{Asteroid} & \colhead{Heliocentric} & \colhead{Geocentric} & \colhead{Phase} & \colhead{H} & \colhead{G} & \colhead{12 $\mu$m} & \colhead{25 $\mu$m}& \colhead{60 $\mu$m} & \colhead{100 $\mu$m} &\colhead{Geometric} &\colhead{Diameter} & \colhead{12 $\mu$m} & \colhead{25 $\mu$m}& \colhead{60 $\mu$m} & \colhead{100 $\mu$m} &\colhead{Geometric} &\colhead{Diameter}  &\colhead{Eta}\\

\colhead{Number} &\colhead{Distance} & \colhead{Distance} & \colhead{angle} & \colhead{} &\colhead{} &\colhead{flux} & \colhead{flux} & \colhead{flux} & \colhead{flux} & \colhead{Albedo} &\colhead{} &\colhead{flux} & \colhead{flux} & \colhead{flux} & \colhead{flux} & \colhead{Albedo} &\colhead{}  & \colhead{}\\

\colhead{} & \colhead{(AU)} & \colhead{(AU)} & \colhead{(deg)} & \colhead{(mag)} & \colhead{} & \colhead{(Jy)} & \colhead{(Jy)} & \colhead{(Jy)} & \colhead{(Jy)} & \colhead{} & \colhead{(km)} & \colhead{(Jy)} & \colhead{(Jy)} & \colhead{(Jy)} & \colhead{(Jy)} & \colhead{} & \colhead{(km)} & \colhead{}\\
}

\startdata
3&2.170&1.700&27.070&5.330&0.320&108.471&173.577&64.128&28.960&0.219&244.172&104.856&177.392&64.721&28.960&0.189&262.797&1.039\\
3&2.124&1.442&24.620&5.330&0.320&104.259&175.596&60.516&30.225&0.194&259.323&101.898&179.372&61.627&30.225&0.191&261.275&0.895\\
4&2.569&2.153&22.310&3.200&0.320&183.566&422.888&129.028&64.33&0.342&520.367&183.566&422.888&129.028&64.330&0.348&515.855&0.842\\
5&2.252&2.111&-26.670&6.850&0.150&17.005&31.496&13.956&4.709&0.313&101.276&16.438&32.188&14.085&4.709&0.167&138.74&1.068\\
5&2.53&2.112&-26.670&6.850&0.150&16.714&29.834&13.630&4.215&0.317&100.658&16.157&30.490&13.756&4.215&0.190&130.021&0.954\\
5&2.252&2.110&-26.670&6.850&0.150&12.772&27.967&9.983&4.166&0.380&91.969&12.347&28.581&10.076&4.166&0.183&132.415&1.216\\
6&2.705&2.578&21.520&5.710&0.240&14.970&38.104&16.491&8.829&0.258&188.766&14.333&38.961&16.802&8.913&0.190&219.839&1.192\\
6&2.691&2.461&21.710&5.710&0.240&18.913&47.775&22.844&20.317&0.230&200.012&18.109&48.848&23.275&10.416&0.165&235.961&1.219\\

\enddata
\tablenotetext{a)}{Iterative estimates using mean temperature color 
color corrections derived from CSAC point source fluxes 
(Table~\ref{table:csac_tab}) corrected for
flux overestimation (see text \S\ref{sec:therm_mods})}.
\end{deluxetable}
\clearpage



\begin{deluxetable}{llclclclclc}
\rotate
\tablecaption{MONTE CARLO STM AND NEATM SOLUTIONS PER
SIGHTING\label{table:mc_sum}}
\tabletypesize{\small}
\setlength{\tabcolsep}{1pt}
\tablewidth{0pt}

\tablehead{
\colhead{Asteriod} & \colhead{STM} & \colhead{STM} & \colhead{STM} &
\colhead{STM} &\colhead{NEATM}&\colhead{NEATM} & \colhead{NEATM } &
\colhead{NEATM } & \colhead{ NEATM} & \colhead{NEATM}\\

\colhead{Number}& \colhead{Albedo} & \colhead{Albedo Error} &
\colhead{Diameter} & \colhead{Diameter Error} & \colhead{Albedo} &
\colhead{Albedo Error} & \colhead{Diameter} & \colhead{Diameter Error} &
\colhead{Eta} &\colhead{ Eta Error}\\

\colhead{}& \colhead{} & \colhead{($\pm$)} & \colhead{(km)} &
\colhead{($\pm$km)} & \colhead{} & \colhead{($\pm$)} & \colhead{(km)} &
\colhead{($\pm$km)}  & \colhead{} &\colhead{($\pm$)}

}

\startdata

       1 &    0.086  &   0.004 &  973.048  & 24.100&     0.097 &    0.004&   91
4.344&    19.653&     0.750&     0.000\\
       1 &    0.095  &   0.004  & 926.012    &19.380&     0.108  &   0.004&   8
70.222&    17.316&     0.750&     0.000\\
       1  &   0.093  &   0.004  & 935.983    &21.892&     0.105   &  0.005&   8
79.717&    21.693&     0.750&     0.000\\

\enddata
\end{deluxetable}
\clearpage


\begin{deluxetable}{lccccccccccccc}
\rotate
\tablecaption{MEAN ALBEDO CATALOG FOR IRAS AND MSX ASTEROIDS\label{table:on_line_examp}}
\tabletypesize{\tiny}
\tablewidth{0pt}

\tablehead{
\colhead{} &\colhead{} & \colhead{Number} & \colhead{} & \colhead{} & \colhead{} & \colhead{} & \colhead{Number} & \colhead{} & \colhead{} & \colhead{} & \colhead{} & \colhead{} & \colhead{} 
\\

\colhead{Asteroid} &\colhead{Source} & \colhead{STM} & \colhead{STM} & \colhead{STM} & \colhead{STM} & \colhead{STM} & \colhead{NEATM} & \colhead{NEATM} & \colhead{NEATM} & \colhead{NEATM} & \colhead{NEATM} & \colhead{NEATM} & \colhead{NEATM} 
\\
\colhead{Number} & \colhead{} & \colhead{Obs.} & \colhead{Albedo} & \colhead{Albedo Error} & \colhead{Diameter} & \colhead{Diameter Error} & \colhead{Obs.} & \colhead{Albedo} & \colhead{Albedo Error} & \colhead{Diameter} & \colhead{Diameter Error} & \colhead{Eta} & \colhead{Eta Error}\\

\colhead{} & \colhead{} & \colhead{} & \colhead{} & \colhead{($\pm$)} & \colhead{(km)} & \colhead{($\pm$km)} & \colhead{} & \colhead{} & \colhead{($\pm$)} & \colhead{(km)} & \colhead{($\pm$km)} & \colhead{} & \colhead{($\pm$)}

}

\startdata
       1 &   IRAS   &   6   &   0.120&     0.019&   855.463&    56.960&       6  &   0.104 &    0.006&   886.476&    27.304&     0.852&     0.102\\
       2   & IRAS    &   7   &  0.175 &    0.015&   479.812&    20.180&       7  &   0.145 &    0.011&   523.982&    20.833&     0.860&     0.050\\
       3   & IRAS     &  8    & 0.213  &   0.012&   248.481&     6.840 &      8  &   0.191  &   0.018&   262.012&    12.047&     0.999&     0.075\\
       4   & IRAS     &  1   &  0.342  &   0.012&   520.367&     6.840&       1 &    0.348   &  0.026&   515.855&    19.247&     0.842&     0.039\\
       5   & IRAS     &  3   &  0.337   &  0.026&    97.968&     3.623&       3  &   0.180    & 0.017&   133.726&     6.468&     1.079&    0.094\\
       6   & IRAS     &  6   &  0.291   &  0.029 &  180.426&     8.497&       6 &    0.203  &   0.021&   214.485&    10.249&     1.114&     0.077\\

\enddata
\end{deluxetable}
\clearpage


\begin{deluxetable}{lcccr}
\tablecaption{ASTEROID DIAMETERS FROM THERMAL MODEL, RADAR, OCCULTATION 
MEASURES\label{table:adtro}}
\tablewidth{0pt}
\tablecolumns{5}
\tablehead{

\colhead{} & \colhead{} & \colhead{} & \colhead{Radar or} & 
\colhead{Source}\\

\colhead{Asteroid} & \colhead{STM} & \colhead{NEATM} & \colhead{ Occultation} & \colhead{Reference\tablenotemark{a}}\\

\colhead{Number} & \colhead{Diameter} & \colhead{Diameter} & \colhead{Diameter} & \colhead{}\\

\colhead{} & \colhead{(km)} & \colhead{(km)} & \colhead{(km)} & \colhead{}

}

\startdata
       1&   855.463$\pm$    56.960&   886.476$\pm$    27.304&   933.750$\pm$     4.789& 1\\
       2&   479.812$\pm$    20.180&   523.982$\pm$    20.833&   539.700$\pm$    21.108 &1\\
       3&   248.481$\pm$     6.840&   262.012$\pm$    12.047&   269.550$\pm$     2.460&1 \\
       4&   520.367$\pm$     6.840&   515.855$\pm$    19.247&   505.000$\pm$     4.252 & 1\\
       5&    97.968$\pm$     3.623&   133.726$\pm$     6.468&   110.500$\pm$     5.423& 1\\
       8&   115.773$\pm$     2.654&   145.755$\pm$     7.060&   162.350$\pm$    11.952& 1\\
              9&   152.410$\pm$     0.008&   178.112$\pm$     0.021&   168.600$\pm$     2.786 &1\\
      13&   223.078$\pm$     3.461&   226.063$\pm$     9.485&   203.400$\pm$     2.970 & 1\\
      17&    80.357$\pm$     3.299&    95.846$\pm$     6.012&    73.000$\pm$     4.072 &1\\
      20&   152.266$\pm$     7.185&   155.040$\pm$     9.617&   145.000$\pm$    17.000 &2\\
      21&    82.671$\pm$     2.568&   104.214$\pm$     7.111&   116.000$\pm$    17.000 &2\\
      23&   101.978$\pm$     3.682&   111.045$\pm$     6.269&   106.000$\pm$    12.000 &3\\
      25&    61.623$\pm$     2.017&    83.388$\pm$     5.846&    79.100$\pm$     1.414 &1\\
      31&   265.939$\pm$     7.001&   286.908$\pm$    12.536&   280.000$\pm$    43.000 & 3\\
      36&   109.093$\pm$     2.852&   121.072$\pm$     9.594&   103.000$\pm$    11.000 & 3\\
      39&   155.123$\pm$     5.652&   184.707$\pm$    10.567&   181.150$\pm$     4.640 &1\\
      41&   172.431$\pm$     4.078&   207.867$\pm$    10.516&   187.300$\pm$     6.220 &1\\
      47&   107.752$\pm$     3.889&   140.814$\pm$     6.580&   138.000$\pm$     1.972&1\\
      51&   155.861$\pm$     3.982&   164.416$\pm$     6.647&   144.300$\pm$     2.746 &1\\
      52&   264.319$\pm$     6.959&   340.748$\pm$    13.370&   319.200$\pm$     4.386 &1\\
      53&   118.637$\pm$     3.549&   115.306$\pm$     5.948&   115.000$\pm$    14.000&3\\
      54&   177.420$\pm$     4.559&   177.676$\pm$     7.551&   147.550$\pm$     1.628 &1\\
      58&    94.553$\pm$     3.113&   102.704$\pm$     7.770&    96.500$\pm$    23.000 &1\\
      60&    63.131$\pm$     3.260&    59.357$\pm$     2.991&    60.000$\pm$     7.000&3\\
          64&    59.461$\pm$     3.927&    83.146$\pm$    26.882&    50.400$\pm$     2.202&1\\
      66&    70.440$\pm$     4.962&    76.578$\pm$     4.826&    69.000$\pm$     9.000&3\\
      70&   105.166$\pm$     2.797&   130.885$\pm$     6.526&   131.300$\pm$     1.082&1\\
      74&   127.668$\pm$     3.339&   132.020$\pm$     5.322&    97.900$\pm$     2.193&1\\
      80&    84.847$\pm$     2.109&    81.149$\pm$     2.652&    79.000$\pm$    10.000&2\\
      83&    79.310$\pm$     3.568&    94.343$\pm$     5.319&    84.000$\pm$     9.000&1\\
      85&   155.238$\pm$     5.089&   168.014$\pm$    10.560&   172.200$\pm$    17.576&1\\
      88&   207.877$\pm$     5.889&   223.199$\pm$    10.219&   207.550$\pm$     2.062&1\\
      89&   134.271$\pm$     4.003&   160.839$\pm$     6.526&   152.950$\pm$     8.424&1\\
      94&   183.167$\pm$     4.610&   206.310$\pm$     8.974&   189.600$\pm$     4.254&1\\
      99&    72.159$\pm$     3.668&    78.009$\pm$     2.557&    71.050$\pm$     1.063&1\\
     101&    57.216$\pm$     1.542&    72.394$\pm$     3.267&    66.000$\pm$     7.000 &3\\
     105&   101.053$\pm$     2.806&   120.572$\pm$     4.743&   104.350$\pm$     3.680&1\\
     106&   127.316$\pm$     3.372&   162.918$\pm$     7.857&   148.600$\pm$    43.600&1\\
     109&    79.486$\pm$     2.420&   111.367$\pm$     9.144&    88.350$\pm$     1.166&1\\
     111&   140.163$\pm$     2.404&   153.170$\pm$     5.790&   131.400$\pm$     1.300&1\\
     114&    81.802$\pm$     2.106&   103.009$\pm$     4.952&   100.000$\pm$    14.000&3\\
     116&    75.953$\pm$     6.952&    84.234$\pm$     6.884&    82.900$\pm$     3.220&1\\
     124&    78.018$\pm$     2.375&    75.021$\pm$     3.370&    65.850$\pm$     2.508&1\\
     134&   122.953$\pm$     4.888&   127.250$\pm$     6.227&   115.050$\pm$    14.320&1\\
     135&    67.288$\pm$     1.886&    92.128$\pm$     5.553&    80.400$\pm$     1.838&1\\
     137&   129.284$\pm$     4.248&   155.643$\pm$    10.656&   144.000$\pm$    16.000&3\\
     141&   110.858$\pm$     3.263&   139.842$\pm$     7.897&   137.600$\pm$    10.610&1\\
     144&   138.402$\pm$     9.118&   161.194$\pm$     9.020&   143.475$\pm$     8.834&1\\
     153&   162.487$\pm$     6.247&   177.424$\pm$     9.525&    81.400$\pm$     7.301&1\\
     182&    36.815$\pm$     1.821&    47.829$\pm$     3.720&    44.000$\pm$    10.000&3\\
     187&   131.420$\pm$     4.659&   132.073$\pm$     7.746&   145.850$\pm$     2.766&1\\
     192&    90.305$\pm$     2.489&   106.881$\pm$     4.972&    98.350$\pm$     2.663&1\\
     198&    56.922$\pm$     3.630&    64.951$\pm$     3.102&    57.000$\pm$     8.000&3\\
     200&   125.108$\pm$     5.903&   135.871$\pm$     6.905&   128.250$\pm$     2.921&1\\
     204&    47.992$\pm$     2.399&    54.150$\pm$     2.411&    50.800$\pm$     1.300&1\\
     208&    42.340$\pm$     3.156&    46.451$\pm$     4.140&    44.400$\pm$     1.924&1\\
     210&    89.045$\pm$     8.350&    85.968$\pm$     3.829&    68.750$\pm$     0.781&1\\
     211&   142.574$\pm$     4.375&   150.951$\pm$     7.501&   143.000$\pm$    16.000&3\\
     216&   119.562$\pm$     3.205&   140.736$\pm$     7.777&   105.950$\pm$     8.964&1\\
     225&   121.595$\pm$     3.341&   122.424$\pm$     5.788&   118.150$\pm$     2.267&1\\
     230&    91.467$\pm$     1.890&   118.601$\pm$     5.246&   101.900$\pm$     8.028&1\\
     238&   149.208$\pm$    14.173&   163.652$\pm$     7.149&   146.750$\pm$    10.132&1\\
     247&   143.994$\pm$     3.901&   156.694$\pm$     6.537&   134.000$\pm$    15.000&3\\
     248&    52.534$\pm$     3.213&    54.208$\pm$     2.683&    54.250$\pm$     1.844&1\\
     266&   112.857$\pm$     2.781&   125.199$\pm$     8.406&   109.000$\pm$    15.000&3\\
     279&   134.096$\pm$     5.365&   136.781$\pm$     7.105&   124.450$\pm$     1.345&1\\
     287&    69.096$\pm$     1.788&    68.111$\pm$     3.154&    75.450$\pm$     2.309&1\\
     306&    48.239$\pm$     1.380&    55.480$\pm$     2.810&    52.250$\pm$     1.118&1\\
     308&   118.450$\pm$     4.131&   162.260$\pm$     6.934&   132.175$\pm$    33.037&1\\
     313&   100.619$\pm$     2.797&    98.365$\pm$     4.106&    96.000$\pm$    14.000&3\\
     324&   248.167$\pm$     8.809&   239.982$\pm$     7.705&   223.300$\pm$     7.839&1\\
     334&   151.454$\pm$    12.849&   182.247$\pm$    19.883&   174.300$\pm$     5.471&1\\
     345&    93.814$\pm$     5.894&   106.199$\pm$     7.809&    99.650$\pm$     0.990&1\\
     350&   109.823$\pm$     4.097&   126.309$\pm$     6.137&    99.800$\pm$     3.421&1\\
     354&   155.338$\pm$     5.161&   165.204$\pm$     7.560&   165.000$\pm$    18.000&3\\
     372&   192.825$\pm$     4.731&   210.109$\pm$     9.500&   195.150$\pm$     4.245&1\\
     381&   120.314$\pm$     3.447&   136.567$\pm$     7.139&   131.550$\pm$     2.025&1\\
     386&   145.672$\pm$     3.771&   186.513$\pm$     9.510&   176.800$\pm$     8.841&1\\
     393&   106.704$\pm$    26.987&   109.777$\pm$    20.808&   125.000$\pm$    20.000&3\\
     404&    98.447$\pm$     4.172&   102.307$\pm$     4.536&   100.200$\pm$     2.309&1\\
     405&   129.605$\pm$     3.970&   134.897$\pm$     6.734&   125.000$\pm$    16.000&3\\
     409&   139.102$\pm$     4.308&   174.535$\pm$     7.395&   165.150$\pm$    33.609&1\\
     411&    65.582$\pm$     2.218&    84.158$\pm$     4.116&    78.000$\pm$     1.556&1\\
     419&   106.639$\pm$     2.944&   139.941$\pm$     5.967&   126.800$\pm$     3.956&1\\
     420&   139.318$\pm$     6.006&   160.326$\pm$     8.801&   145.800$\pm$     1.924&1\\
     429&    61.054$\pm$     1.737&    73.809$\pm$     3.548&    70.000$\pm$    10.000&3\\
     431&    84.580$\pm$     2.590&   102.113$\pm$     5.015&    69.200$\pm$     5.883&!\\
     444&   159.343$\pm$    13.673&   167.732$\pm$     6.772&   169.150$\pm$    14.348&1\\
     466&   114.911$\pm$     3.557&   131.924$\pm$     5.654&   126.950$\pm$     4.000&1\\
     471&   121.829$\pm$     4.641&   138.542$\pm$     6.229&   130.900$\pm$     2.012&1\\
     476&   124.609$\pm$     4.018&   124.022$\pm$     5.835&   102.800$\pm$    10.265&1\\
     488&   155.986$\pm$     6.944&   161.588$\pm$     7.390&   150.000$\pm$    21.000&3\\
     498&    79.162$\pm$     3.536&    90.904$\pm$     4.181&    79.300$\pm$     2.961&1\\
     522&   104.419$\pm$     7.747&   107.284$\pm$     9.370&    89.300$\pm$     3.467&1\\
     526&    41.241$\pm$     7.747&    46.173$\pm$     1.727&    45.850$\pm$     1.030&1\\
     530&    77.184$\pm$     2.698&    89.082$\pm$     5.887&    88.250$\pm$     3.183&1\\
     566&   171.473$\pm$     1.191&   190.084$\pm$     7.909&   137.350$\pm$    13.741&1\\
     568&    84.541$\pm$     4.290&    92.405$\pm$     4.763&    76.250$\pm$     3.585&1\\
     578&    73.227$\pm$     2.549&    72.777$\pm$     3.293&    72.875$\pm$     6.395&1\\
     580&    42.101$\pm$     2.036&    53.520$\pm$     2.705&    49.800$\pm$     0.849&1\\
     693&    58.296$\pm$     1.784&    77.257$\pm$     3.449&    81.400$\pm$     1.200&1\\
     697&    70.259$\pm$     2.610&    86.817$\pm$     3.948&    75.300$\pm$     2.377&1\\
     702&   169.078$\pm$     4.344&   215.635$\pm$     9.235&   203.300$\pm$     4.188&1\\
     704&   285.197$\pm$     6.999&   358.470$\pm$    14.745&   329.950$\pm$    11.272&1\\
     712&   134.508$\pm$     3.720&   141.183$\pm$     6.971&   118.950$\pm$     2.476&1\\
     747&   159.136$\pm$     5.164&   185.980$\pm$    10.415&   182.450$\pm$    11.985&1\\
     757&    26.810$\pm$     1.222&    39.395$\pm$     3.536&    36.850$\pm$     1.432&1\\
     790&   165.446$\pm$     9.333&   173.412$\pm$     7.380&   153.350$\pm$     2.773&1\\
     791&    93.427$\pm$     2.834&   115.038$\pm$     4.663&    84.600$\pm$     4.386&1\\
     828&    47.734$\pm$     1.798&    59.854$\pm$     3.655&    51.500$\pm$     2.921&1\\
     903&    67.620$\pm$     3.070&    69.855$\pm$     3.132&    81.400$\pm$     1.200&1\\
     914&    66.960$\pm$     2.103&    83.642$\pm$     3.762&    91.450$\pm$     2.184&1\\
     925&    47.856$\pm$     1.551&    63.805$\pm$     2.913&    61.850$\pm$     1.887&1\\
     976&    82.830$\pm$     4.103&    86.958$\pm$     5.074&    67.400$\pm$    10.474&1\\
    1263&    51.232$\pm$     2.047&    54.397$\pm$     2.722&    45.050$\pm$     1.166&1\\
    1409&    36.294$\pm$     1.694&    41.862$\pm$     2.747&    81.400$\pm$     4.817&1\\
    1512&    79.277$\pm$     3.628&    88.479$\pm$     4.289&    67.500$\pm$     2.657&1\\
    1867&   111.352$\pm$     4.666&   137.194$\pm$     3.874&   127.300$\pm$     5.859&1\\
  
\enddata
\tablenotetext{a)}{Reference values correspond to the following papers: 1=\citet{dh09}, 2=\citet{magri99}, 3=\citet{mag07}}
\end{deluxetable}


\begin{deluxetable}{lcccr}
\tablecaption{THERMAL MODEL DIAMETERS FROM STM, NEATM, and 
TPM\tablenotemark{a}\label{table:sn_v_delbo}}
\tablewidth{0pt}
\tablecolumns{5}

\tablehead{
\colhead{Asteroid} & \colhead{Asteroid} & \colhead{TPM\tablenotemark{a}} 
& \colhead{STM} & \colhead{NEATM}\\

\colhead{Number} & \colhead{Name} & \colhead{Diameter} & \colhead{Diameter} &
\colhead{Diameter}\\

\colhead{} & \colhead{} & \colhead{(km)} & \colhead{(km)} &\colhead{(km)}

}

\startdata

21& Lutetia& 107-114& 82.67 $\pm$ 2.57&104.21 $\pm$7.11  \\ 

32& Pomona& 84-86&   80.73 $\pm$ 1.99  &82.94 $\pm$4.08     \\

44& Nysa& 80-82      &     71.95 $\pm$2.14     &   76.50 $\pm$3.83  \\

110& Lydia& 90-97       & 73.30 $\pm$2.40 &      90.33 $\pm$5.47\\

115& Thyra& 90-94      & 66.70 $\pm$1.66 &        83.02 $\pm$4.86\\

277& Elvira& 36-40      & 27.77 $\pm$ 1.24 &       33.98 $\pm$1.97\\

306& Unitas& 55-57     & 48.24 $\pm$1.38&        55.48 $\pm$2.81\\ 

382& Dodona& 74-76  & 49.02 $\pm$1.83 &        68.66 $\pm$3.86\\

694& Ekard& 108-111 & 90.51 $\pm$ 4.34 &     101.87 $\pm$17.15\\

720& Bohlinia& 40-42 & 28.61 $\pm$ 0.98 &        40.21 $\pm$3.13\\

\enddata
\tablenotetext{a)}{Thermal Physical Model of \citet{delbtan09}.}
\end{deluxetable}
\clearpage


\begin{deluxetable}{lc|cc|cr}
\tablecaption{ALBEDO VALUES FOR DEMEO TAXONOMIC TYPES\label{table:demotax}}
\tablewidth{0pt}
\tablecolumns{6}

\tablehead{
\colhead{Type} & 
\colhead{Number Objects} & 
\colhead{STM} & 
\colhead{STM} & 
\colhead{NEATM} & 
\colhead{NEATM}\\
\colhead{} & 
\colhead{of Type} & 
\colhead{Albedo} & 
\colhead{Albedo Range} & 
\colhead{Albedo} & 
\colhead{Albedo Range}
}

\startdata
\cutinhead{S complex}\\
S & 43&0.260&     0.071&0.174&     0.039\\
Sa & 1&   0.397 &&0.339&\\
Sq &4&     0.258&     0.031&0.213&     0.064\\
Sr & 2& 0.219&     0.029&0.167&    0.029\\

\cutinhead{C complex}\\
B & 2&0.265&      0.128&0.083&     0.087\\
C &12&0.070&    0.023&0.053&     0.018\\
Cb &3&0.074&     0.047&0.063&     0.039\\
Cg &1&0.105&&0.058&\\
Cgh &10& 0.138&      0.144&0.103&     0.083\\
Ch &16& 0.069&     0.014&0.052&    0.009\\
\\
\\
\cutinhead{X complex}\\
X &3& 0.100&     0.066&0.077&   0.054\\
Xc &3& 0.200&      0.121&0.137&     0.065\\
Xe &3& 0.159&    0.009&0.128&   0.019\\
Xk &16& 0.177&      0.102&0.113&     0.054\\

\cutinhead{End Members}\\
D &11& 0.082&     0.031&0.067&     0.029\\
K  & 12&0.163&     0.073&0.116&    0.042\\
L  &12& 0.141&     0.037&0.110&     0.035\\
T &4& 0.054&     0.013&0.042&     0.005\\
A &4& 0.265&0.117&0.179&     0.035\\
R &1& 0.507&&0.411&\\
V &1& 0.3742&&0.348&\\

\enddata
\end{deluxetable}

%
%

\clearpage


\clearpage


\clearpage


\clearpage


\clearpage


\clearpage


\clearpage


\clearpage

\end{document}